\documentstyle [12pt,twoside]{article}
  
\begin{document}
\begin{flushright}
UBCTP-93-016
\end{flushright}
\vspace*{1cm}
\centerline{\large{\bf Early Direct Detection}}
\centerline{\large{\bf of}}
\centerline{\large{\bf Gravity Waves}}
\vspace*{1cm}
\centerline{\bf Redouane Fakir}
\vspace*{0.5cm}
\centerline{\em Cosmology Group, Department of Physics}
\centerline{\em University of British Columbia}
\centerline{\em 6224 Agriculture Road}
\centerline{\em Vancouver, B. C. V6T 1Z1}
\centerline{\em CANADA}
\vspace*{1.cm}
\centerline{\bf Abstract}
\vspace*{0.5cm}
%\addtolength{\baselineskip}{\baselineskip}

Recently, the possibility has emerged of an early
detection of astrophysical gravity waves.
In certain astronomical configurations, and
through a new light-deflection effect, gravity waves
can cause apparent shifts in stellar angular positions as
large as $10^{-7}arcsec$. In these same configurations,
the magnitude of the gravity-wave-induced time-delay effect
can exceed $10^{-14}$. Both these figures
lie just at present-day theoretical limits of detectability.
For instance, cases are described where the very faint
neutron-star gravity waves could soon become detectable.
The detection meant here involves direct observations
of the very wave-forms.

\clearpage

Perhaps the earliest understood physical effect of
gravity waves, is their modulating of proper distances [1-4].
The first bar detectors \cite{weber}, as well
as recent detection projects such as LIGO
\cite{ligo}
 and VIRGO \cite{virgo}, are based on that effect.

The experimental challenge facing such detection
efforts is daunting. The expected distance modulations
have about the same magnitude as the gravity-wave's
amplitude, which is typically smaller than $10^{-22}$
in the vicinity of the Earth. Thus, these experimentsonics an
involve detecting shifts much smaller than one Fermi
in distances of the order of a kilometer.

Not long ago, it was proposed to explore an approach
to gravity-wave detection based on accelerations of
null, rather than timelike geodesics \cite{fakir1}.

The simplest illustration of this idea is the shifting
of apparent stellar positions due to an intervening
gravitational pulse \cite{fakir2}. Suppose a supernova
flash hits the Earth, coming from the northern
celestial hemisphere. This is
an indication that a gravitational pulse has also
just whipped passed the Earth, and is now interposed
between us and all the southern celestial hemisphere.
It was calculated that the angular positions of
southern stars would then experience apparent shifts
of the order of the pulse's amplitude $h$:
\newline\begin{equation}
|\delta\alpha| \approx {1\over 2} h \sin\alpha  \  \ ,
\end{equation}\newline
where $\alpha$ is the angle of incidence of the light
rays with respect to the gravitational wave-front.
In the case of ``pulses with memory''[10-13], such shifts
can be quasi-permanent.

Quantitatively, this version of the effect does nothing
to improve the prospects of gravity-wave detection.
The angular shifts resulting from eq.(1) should be
smaller than $10^{-17}arcsec$, while the present
theoretical limit for angular resolutions from
Very-Long-Baseline Interferometry is about $10^{-7}arcsec$.

However, this effect presents a feature that distinguishes
it qualitatively from many others. In eq.(1), $h$ is not
the amplitude of the waves when they meet the Earth.
Instead, $h$ is the waves' amplitude when they meet the
stellar photons, which eventually reach the Earth.
This amounts to a prospect of {\em remote probing of
gravity waves}.

Since $h<h_{(Earth)}$ in the case above, this feature can only
worsen the observational situation for this particular
illustration. Nevertheless, this same feature can dramatically
improve the detectability of the effect, if the configuration
defined by the gravity-wave source, the Earth and the light
source is, in a sense, inverted \cite{fakir3,fakir4}.
 In the new configurations,
probing the waves at a distance could mean probing them
in regions of space where $h$ is larger, not smaller,
than $h_{(Earth)}$.

In the previous illustration, the Earth was placed between
the gravity-wave source and the light source. Consider, now,
a situation where it is the gravity-wave source that is
placed between the Earth and the light source.
Then the photons, during their journey towards the Earth,
would have encountered gravity-wave crests with heights
ranging from $h_{(light-source)}$ to $h(b)$
to $h_{(Earth)}$, where $b$ is the distance of closest approach
between the photons and the gravity-wave source,
the ``impact parameter.''

The hope, of course, is that the photon will ``remember''
the highest amplitude of gravity waves it sees on its
way to the Earth. If so, the analogue of eq.(1) for
the new configuration would exhibit $h(b)$ on the
right-hand side, not the much smaller $h_{(Earth)}$ or
$h_{(light-source)}$. The whole scheme would then amount
to remote probing of {\em strong} gravity-wave sites.

A priori, there are several reasons to fear this scheme
would not work. The physics of the photon's encounter
with gravity-waves is more involved in this latter case
of spherical wave fronts than in the former plane-wave
case.

For example, one could question whether the deflections acquired
 by a photon during
the ``ingoing'' phase (approaching the gravity-wave source)
are not cancelled by deflections during the outgoing phase.
Fortunately, the calculation shows this is not the case.

One could also wonder if there would be deflections at all
during the outgoing phase. The gravity-wave crests travel
at the same speed as the photon itself. Now, the photon
is only sensitive to {\em variations} in $h$, and it would
see no such variations if it travels along with the gravity
waves. Nevertheless, in most actual situations, photons
and gravity-wave fronts travel {\em at an angle}. Hence,
in the outgoing phase also, photons
may see changes in $h$ and experience deflections.

Let the gravity-wave mode of interest be described by
\newline\begin{equation}
h = {H\over r}\exp \{i\Omega (r-t+t_{ph})\} \ \ ,
\end{equation}\newline
where $t_{ph}$ determines the wave's phase. $H$ is
a constant that encodes the intrinsic strength of the
source.
Working in a spherical transverse-traceless gauge,
projecting the problem onto a plane containing the
Earth and the light and gravity-wave sources, and considering
the optimal alignment case where $b\Omega<1$,
 one finds \cite{fakir3}
\newline\begin{equation}
|\delta\phi|_{optimal} \approx  {3\over 4} \pi \Omega H
= {3\over 2} \pi^{2} | h(r=\Lambda) | \  \ ,
\end{equation}\newline
where $\Lambda$ is one gravitational wavelength. The angle
$\phi$ is close to $0$, $\pi/2$ and $\pi$ at the light source,
the gravity-wave source and the Earth, respectively.

Let us generalize this result to arbitrary values of the
impact parameter $b$. We can infer from eq.(16) of the
above reference that
\newline\[
\delta\phi \approx {H\over b} e^{i\Omega t_{ph}}
\int_{0}^{\pi} d\phi
\exp\left\{ i\Omega b {1+\cos\phi\over \sin\phi} \right\}
\]
\newline\begin{equation}
\times \left[ \sin\phi - {3\over 2} \sin^{3}\phi
+ i\Omega b \left( {\sin^{2}\phi\over 2} - 1 - \cos\phi \right)
\right]  \  \ .
\end{equation}\newline
(This was obtained by comparing the two ends of the
trajectory $\phi\approx 0$ and $\phi\approx \pi$. One can
show that $\delta\phi = b [ u_{1}(\phi\approx 0) +
 u_{1}(\phi\approx\pi) ]$ , where $u_{1}$ is the fluctuation
of $1/r$.)

Eq.(4) can be rewritten as
\newline\begin{equation}
\delta\phi \approx {H\over b} e^{i\Omega t_{ph}}
\int_{0}^{\infty}
dx {4x e^{ib\Omega x}\over (1+x^{2})^{2}}
\left[ 1 - {6x^{2}\over (1+x^{2})^{2}}
- {ibx^{3}\over 1+x^{2}} \right] \  \ ,
\end{equation}\newline
which integrates nicely to the analytical formula
\newline\begin{equation}
\delta\phi \approx
{1\over 2} H \Omega e^{i\Omega t_{ph}}
\left[  (b\Omega +1) e^{b\Omega} E_{1}(b\Omega) +
(b\Omega-1) e^{-b\Omega} E_{1}(-b\Omega)  \right]  \  \ .
\end{equation}\newline
$E_{1}(z)$ is the exponential integral function
\newline\begin{equation}
E_{1}(z) = \int_{1}^{\infty} {e^{-zt}\over t} dt \ \ \ \
 Re(z)>0 \  \ ,
\end{equation}\newline
extended analytically to the entire complexe plane except $z=0$.
It is straightforward to verify that eq.(5) integrates to eq.(3)
in the limit $b\Omega<<1$.

Thus,
the gravity-wave-induced deflection is equal to
the wave's amplitude at only one gravitational wavelength
from the source, times a factor that decreases slightly
faster than $1/b\Omega$.

There are several actual astronomical
configurations to which this approach can be applied.
The candidates fall into two classes. In the first,
the gravity-wave source and the light source are aligned
with the Earth by pure chance. They are two unrelated, far apart
celestial objects. Because of the huge number of binary stars
in the Galaxy, also because of their relatively large gravity-wave
amplitude and wavelength, a lucky alignment of a binary star
with some more distant light source would be the typical
candidate in this class. Neutron stars are too scarce,
too weak, and have too short wavelengths to qualify in this
context.

Numerically, candidates in this first category could
produce optimal shifts of about $10^{-8}arcsec$,
remarkably close to the $10^{-7}arcsec$ theoretical
limit of observability. For
instance, a binary source with $H = 6$cm and an orbital
period of $10$ hours (i.e. the gravity-wave period is
$2\pi/\Omega = 5$ hours) would produce shifts of
$2\times 10^{-8}arcsec$.

For the second class of candidates, the gravity-wave source
and the light source are locked into tight gravitationally
bound systems. Common examples of this in the galaxy are
stars (as light sources) and binaries (as gravity-wave sources)
locked into multiple-star systems or even globular clusters.
Of particular observational importance is the case of
a binary formed by a neutron star (as the gravity-wave source)
and some companion star (as the light source).

Comparison of typical gravitational wavelengths and
typical separations shows that the
alignment requirement, for this category, is satisfied
naturally \cite{fakir3}. Interestingly, through this
mechanism, the very faint
gravity waves from neutron stars might be no less observable
than  the ones from binary stars or from cataclysmic sources.

Take, for instance, a system like the well studied binary pulsar
PSR B1913+16 \cite{taylor}. If we are to believe
the above analysis, this system and alike could be, indeed,
very promising sites
for direct gravity-wave detection. (This is, of course, besides the
indirect evidence for the existence of gravity waves already provided
by the observed secular slow-down of this binary pulsar.)

Consider first, as the gravity-wave source, the dark neutron
star that revolves around the actual $17$Hz pulsar.
Once every $7^{h}45'$, the two stars come to within only
one light-second (about half a solar radius) of each-other.
This is, at most, of the same order of magnitude as the darker
companion's gravitational wavelength.

Numerically, the shifts produced in this case
may well reach the $10^{-7}arcsec$ detectability limit
\cite{fakir3}.
For instance, if the companion is a $1$Hz neutron star
(i.e. the gravitational frequency is $\Omega/2\pi = 2 sec^{-1}$),
radiating, perhaps through the Chandrasekhar-
Friedman-Schutz mechanism \cite{cfs}, with a strength
$H = 10^{-6}$m, then the shifts are  typically $5\times 10^{-8}arcsec$.

Besides the future prospects of achieving angular resolutions
of the order of $10^{-7}arcsec$ for radio sources by space-based
 interferometry,
there has been considerable progress, recently, towards reaching
very high angular resolutions for optical sources as well
\cite{hipparcos,roemer}.
Also, the increase in angular resolution power has been accompanied
by a considerable improvement in photometric sensitivities, potentially
revealing a number of new stellar systems that could
be relevant to this study.

In principle, there are two more sources of gravity waves that could
be affecting the apparent position of that same light source.
One is the pulsar itself. Being a neutron star that rotates
$17$ times per second, it should be emitting gravity waves
at a frequency of $34$Hz. However, 1) the angle between the
electromagnetic and the gravitational directions of propagation
is very small in this case, 2) here there is no incoming, only
an outgoing phase. As mentioned above, the combination
of these two facts means that the light from the pulsar is
unlikely to be deflected by the pulsar's own gravity waves.

The other additional source of waves is the binary system as a
whole. (These are the waves for which there is
already indirect observational evidence.) Here also,
the shortness of the incoming phase and the smallness
of the angle between the electromagnetic and the gravitational
directions of propagation are a concern. More importantly,
there are more considerations that have to be taken into account,
before one can make predictions in this case. The photons, here,
originate from the gravity-wave source itself, and traverse
the near-zone ($r<\Lambda$) before reaching the radiation zone,
where our calculations are valid. Such cases necessitate a separate
study, where, in particular, dipole Neutonian contributions to
the deflection would have to be included.

The consideration of pulsars as sources of the deflected light
quickly lead to another gravity-wave detection prospect.
Following the above study, the next logical step is to try
to exploit the exceptional properties of pulsars, especially
the high stability of their period. This
quickly lead to another gravity-wave detection prospect.
It stems from the application of an effect that has little
to do with the gravity-wave-induced light deflection,
namely the gravity-wave-induced {\em modulation of time delays},
to the same configurations discussed above \cite{fakir4}.

The possibility of gravity-wave detection through the
modulation of pulsar frequencies by {\em plane} waves
has already been extensively explored \cite{pulsars,taylor}.
The experimental effort in this field has made it
possible to detect fractional frequency modulations as faint as
$10^{-14}$ Thus,
stringent upper limits could be imposed on
the cosmological and the galactic
gravity-wave backgrounds. Recently, we also learned that,
in the wake of this effort, the contribution of individual
binary stars was also considered in one instance \cite{sazhin}.
Unfortunately, this initial, heuristic investigation was not
followed up by the
consideration of more promising candidates, such as systems
consisting of
gravitationally bound light and gravity-wave sources.

In complete analogy with our discussion of the
gravity-wave-induced light-deflection effect,
the hope here is that, 1) it can be
shown rigorously that the crossing of a zone of spherical gravity waves
does result in a net frequency modulation;
2) the strongest gravity-wave amplitudes encountered along
the trajectory do contribute to the net modulation.

The same worries we had initially for the working
of the light-deflection effect (see above), can be expressed
here.
Once again, the calculation shows these worries not to
be founded \cite{fakir4}.
 It was shown
that spherical gravity waves can induce time-delay fluctuations
$\delta(\Delta t)$ that vary at a rate
\newline\begin{equation}
{d\ \over dt_{ph}} \delta(\Delta t)  \approx
{1\over 2} \Omega H  e^{i\Omega t_{ph}}
\int_{\phi_{initial}}^{\phi_{final}}d\phi
\sin\phi \exp\left\{ ib\Omega{1+\cos\phi\over \sin\phi} \right\}
 \  \ .
\end{equation}\newline
($\Delta t$ is the total time it takes a photon to travel
from the light source to the Earth, via the gravity-wave
source.)

Here also, the problem has a compact analytical solution.
A change of variables can put eq.(8) in the form
\newline\begin{equation}
\left| {d\ \over dt_{ph}} \delta(\Delta t) \right| \approx
2 \Omega H \left|
\int_{0}^{\infty} {x\over (1+x^{2})^{2}} e^{i b \Omega x} dx
\right| \  \ .
\end{equation}\newline
This integrates to
\newline\begin{equation}
\left| {d\ \over dt_{ph}} \delta(\Delta t) \right| \approx
H \Omega \left| {b\Omega\over 2}
\left( e^{b\Omega} E_{1}(b\Omega) -
e^{-b\Omega} E_{1}(-b\Omega) \right) - 1 \right| \  \ .
\end{equation}\newline

Hence, numerically, this second effect behaves just like
the first one
(eq.(6)),
at least in orders of magnitude: For optimal alignments, it
is as high as the waves' amplitude only one gravitational
wavelength away from the source. For larger impact parameters,
the effect decreases roughly like $1/b\Omega$.

To use the same numerical illustration as for the previous effect,
a binary star with $H = 6$cm and a gravity-wave period (half the
orbital period) $T = 2\pi/\Omega = 5$ hours, would produce fractional
frequency modulations of about $4\times 10^{-14}$. A neutron star
with $H = 10^{-6}$m and $T = 2\pi/\Omega = 0.5$sec, would yield frequency
modulations as strong as $9\times 10^{-14}$.

Retrospectively, this approach to gravity-wave detection has exploited
a perhaps curious observational fact. For several cases of gravity-wave
sources that are members of gravitationally bound stellar systems,
the stellar separations can be as small as only one gravitational
wavelength or so. Thus, in a dense globular cluster, the average
stellar separation is of the same order of magnitude as the
gravitational wavelength of a typical binary star. For a binary system,
one member of which is a neutron star, the orbital size can be
comparable to that neutron-star's gravitational wavelength.
Hence, there exists many astronomical sites where light sources are
constantly moving close to, or even within gravity-wave near-zones.

\vspace*{2.cm}
\centerline{\bf Acknowledgements}
\vspace*{0.5cm}
Throughout this research, I benefitted from the precious
counsel and support of my teacher, W.G. Unruh. I would like also
to thank S. Braham for several informative discussions.

This work would not have been possible without the sustained
material support of the Cosmology Group in the Departement of Physics,
University of British Columbia.

\end{document}